\documentclass[%
 reprint,
superscriptaddress,
 amsmath,amssymb,
 aps,
pra,
]{revtex4-1}

\usepackage{graphicx}
\usepackage{dcolumn}
\usepackage{bm}
\usepackage{epstopdf}
\usepackage{caption}

\def\ket#1{\mathinner{|{#1}\rangle}}

\begin{document}

\preprint{APS/123-QED}
\title{Stability of a trapped atom clock on a chip}
\author{R. Szmuk}
\affiliation{LNE-SYRTE, Observatoire de Paris, PSL Research University, CNRS, Sorbonne Universit\'es, UPMC Univ. Paris 06, 61 avenue de l'Observatoire, 75014 Paris, France}
\author{V. Dugrain}
\affiliation{Laboratoire Kastler Brossel, ENS, UPMC, CNRS, 24 rue Lhomond, 75005 Paris, France}
\author{W. Maineult}
\affiliation{LNE-SYRTE, Observatoire de Paris, PSL Research University, CNRS, Sorbonne Universit\'es, UPMC Univ. Paris 06, 61 avenue de l'Observatoire, 75014 Paris, France}
\author{J. Reichel}
\affiliation{Laboratoire Kastler Brossel, ENS, UPMC, CNRS, 24 rue Lhomond, 75005 Paris, France}
\author{P. Rosenbusch}
\email{Peter.Rosenbusch@obspm.fr}
\affiliation{LNE-SYRTE, Observatoire de Paris, PSL Research University, CNRS, Sorbonne Universit\'es, UPMC Univ. Paris 06, 61 avenue de l'Observatoire, 75014 Paris, France}


\date{\today}

\begin{abstract}
We present a compact atomic clock interrogating ultracold $^{87}$Rb magnetically trapped on an atom chip. Very long coherence times sustained by spin self-rephasing allow us to interrogate
the atomic transition with 85\% contrast at 5 s Ramsey time. The clock exhibits a fractional frequency stability of $5.8\times 10^{-13}$ at 1 s and is likely to integrate into the $10^{-15}$ range in less than a day. A detailed analysis of 7 noise sources explains the measured frequency stability. Fluctuations in the atom temperature (0.4~nK shot-to-shot) and in the offset magnetic field ($5\times10^{-6}$ relative fluctuations shot-to-shot) are the main noise sources together with the local oscillator, which is degraded by the 30\% duty cycle. The analysis suggests technical improvements to be implemented in a future second generation set-up
. 
The results demonstrate the remarkable degree of technical control that can be reached in an atom chip experiment.

\end{abstract}

\pacs{Valid PACS appear here}
\maketitle


\section{introduction}
Atomic clocks are behind many everyday tasks and numerous fundamental science tests. 
Their performance  has made a big leap through the discovery of laser cooling \cite{cohen1998manipulating,chu1998manipulation,phillips1998laser} giving the ability to control the atom position on the mm scale. It has led to the development of atomic fountain clocks \cite{kasevich1989rf,clairon1991ramsey} which 
have reached a stability limited only by fundamental physics properties, i.e. quantum projection noise 
and Fourier-limited linewidth \cite{santarelli1999quantum}.
 While these laboratory-size set-ups are today's primary standards, mobile applications such as telecommunication, satellite-aided navigation \cite{dow2009international} or spacecraft navigation \cite{ely2013deep} call for  smaller instruments with litre-scale volume.
In this context, it is natural to consider trapped atoms. 
The trap overcomes gravity and thermal expansion and thereby enables further gain on the interrogation time. It makes interrogation time independent of apparatus size. 
Typical storage times of neutral atoms range from a few seconds to minutes \cite{ott2001bose,harber2003thermally}. 
Thus a trapped atom clock with long interrogation times could measure energy differences in the mHz range in one single shot.   Hence, if trap-induced fluctuations can be kept low,  trapped atoms could not only define time with this resolution, but could also be adapted to measure other physical quantities like electromagnetic fields, 
 accelerations 
  or rotations 
  with very high sensitivity. 
A founding step towards very long interrogation  of trapped neutral atoms was made in our group through the discovery of spin self-rephasing \cite{deutsch2010spin} which sustains several tens of seconds coherence time \cite{deutsch2010spin,kleinebuning2011extended,bernon2013manipulation}. This  rivals trapped ion clocks, the best of which has shown  65~s interrogation time and a stability of $2~10^{-14}$ at 1~s \cite{burt2008compensated,tjoelker1996mercury}. 
 It is to be compared to compact clocks using thermal vapour and buffer gas \cite{kang2015demonstration,micalizio2012metrological,danet2014dick} or laser cooled atoms \cite{esnault2011horace, muller2011compact,vshah2012,donley2014frequency}. Among these the record stability is $1.6\times 10^{-13}$ at 1~s  \cite{micalizio2012metrological}. 
Clocks with neutral atoms trapped in an optical lattice have reached impressive stabilities down to the $10^{-18}$ range \cite{bloom2014optical,hinkley2013atomic} but their interrogation time is so far limited by the local oscillator. Research into making such clocks transportable is on-going \cite{poli2014transportable,bongs2015development}. 
We describe the realization of a compact clock  using  neutral atoms trapped on an atom chip and analyze trap-induced fluctuations.

Our "trapped atom clock on a chip" (TACC) employs laser cooling and evaporative cooling to reach ultra-cold temperatures  where neutral atoms can be held in a magnetic trap. Realising a 5~s Ramsey time, we obtain  100~mHz linewidth and 85\% contrast on the hyperfine transition of  $^{87}$Rb. We measure the fractional frequency stability as
 $5.8\times10^{-13}\tau^{-1/2}$. It is reproduced by analyzing several noise contributions, in particular atom number, temperature and magnetic field fluctuations.
 The compact set-up is realized through the atom chip technology \cite{reichel2011atom}, which builds on the vast knowledge of micro-fabrication. 
 The use of atom chips is also widespread for the study of Bose-Einstein condensates \cite{hansel2001bose,ott2001bose}, degenerate Fermi gases \cite{aubin2006rapid} and gases in low dimensions \cite{esteve2006observations,hofferberth2007non}. Other experiments strive for the realization of quantum information processors \cite{schmiedmayer2002quantum, treutlein2006microwave, leung2011microtrap}. The high sensitivity and micron-scale position control have been used for  probing  static magnetic \cite{wildermuth2005bose} and electric \cite{tauschinsky2010spatially} fields as well as microwaves  \cite{ockeloen2013quantum}. 
 Creating atom interferometers \cite{cronin2009optics} on atom chips 
is equally appealing. 
Here, an on-chip high stability atomic clock not only provides an excellent candidate for mobile timing applications, it  also takes a pioneering role among this broad range of atom chip experiments, demonstrating that experimental parameters can be mastered to the fundamental physics limit.

This paper is organised as follows: we first describe the atomic levels and the experimental set-up. Then we give the evaluation of the clock stability and an analysis of all major noise sources.

\section{Atomic levels}

\begin{figure}[h!]
\centering
\includegraphics[width=0.7 \columnwidth]{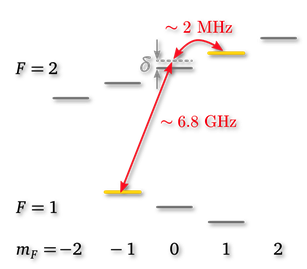}
\caption{Level scheme of the $^{87}$Rb ground state. The two magnetically trappable clock states $\ket{F=1, m_F=-1}$ and $\ket{F=2, m_F=1}$ are coupled via a 2-photon, microwave and radiofrequency transition, where the microwave is tuned $500~\text{kHz}$ above the $\ket{1,-1}\to\ket{2,0}$ transition.}
\label{fig:clock_transition}
\end{figure}

We 
interrogate the hyperfine transition of $^{87}$Rb (figure \ref{fig:clock_transition}). A two photon drive couples the magnetically trappable states $\ket{1}\equiv\ket{F=1,m_F=-1}$ and $\ket{2}\equiv\ket{F=2,m_F=1}$, whose transition frequency exhibits a minimum at a magnetic field near $B_m\approx3.229$~G \cite{matthews1999watching,harber2002effect}. 
This $\text{2}^{\text{nd}}$ order dependence strongly reduces the clock frequency sensitivity to magnetic field fluctuations. It assures that atoms with different trajectories within the trap still experience similar Zeeman shifts. Furthermore, by tuning the offset  magnetic field, the inhomogeneity from the negative collisional shift \cite{harber2002effect} 
  can be compensated to give a quasi position-invariant overall shift \cite{rosenbusch2009magnetically}. Under these conditions of strongly reduced inhomogeneity we have shown that spin self-rephasing can overcome dephasing and that coherence times of $58\pm12$~s \cite{deutsch2010spin} can be reached. It confirms the possibility to create a high stability clock \cite{treutlein2004coherence}.  

\section{Experimental set-up}
\label{sectionSetup}

\begin{figure}[h!]
\centering
\includegraphics[trim = 0mm 0mm 0mm 0mm, clip,width=\columnwidth]{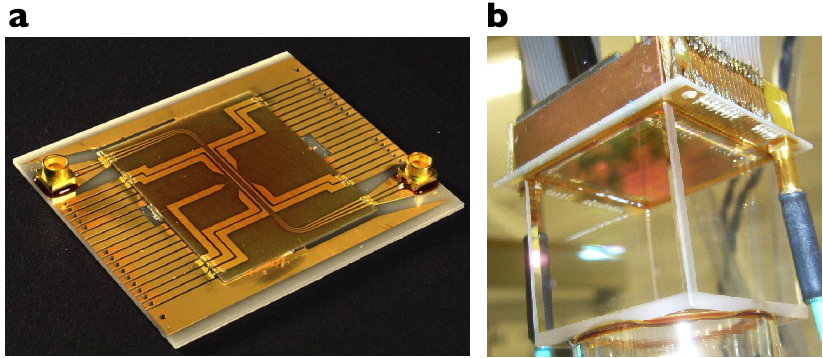}
\caption{(Color online) (a) The implemented atom chip. One identifies the Z-shaped coplanar waveguide which serves for atom trapping and transport of the microwave interrogation signal. The outer dimensions are $38 \times 45.5~\text{mm}^2$. (b)
The chip constitutes one of the facets of the vacuum cell facilitating electrical contact. 
The cell is surrounded by a $10\times10\times15~\text{cm}^3$ cage of Helmholtz coils and a 30 cm diameter optical table
holding all optical beam expanders. The cell is evacuated via standard UHV equipment.}
\label{fig:TACC_chip_cell}
\end{figure}

The experimental set-up, details of which are given in \cite{lacroute2010preliminary}, is similar to compact atom chip experiments reported previously \cite{reichel1999atomic,  bohi2010imaging}. All experimental steps, laser cooling, evaporative cooling, interrogation and detection take place in a $\sim(5$~cm$)^3$ glass cell where one cell wall is replaced by the atom chip   (figure \ref{fig:TACC_chip_cell}). In this first-generation set-up, a 25~l/s ion pump is connected via standard vacuum components. It evacuates the cell to a pressure of $\sim1\times10^{-9}$~mbar. The cell is surrounded by a $10\times10\times15~\text{cm}^3$ cage of Helmholtz coils. A 30~cm diameter optical table holds the coil cage as well as all beam expanders necessary for cooling and detection and is surrounded by two layers of magnetic shielding. 

The timing sequence (table \ref{timingSequence}) starts with a mirror MOT \cite{reichel1999atomic} loading $\sim3\times10^6$ atoms in $\sim7$~s from the background vapor. The MOT magnetic field is generated by one of the coils  and a U-shaped copper structure placed behind the atom chip \cite{wildermuth2004optimized}. Compressing the MOT followed by  5~ms  optical molasses cools the atoms to $\sim20~\mu$K. The cloud is then optically pumped to the $\ket{1}$ state and transferred to the magnetic trap. It is gradually compressed to perform RF evaporation, which takes $\sim3$~s. A $0.7$~s decompression ramp transfers the atoms to the final interrogation trap with trap frequencies $(\omega_x,\omega_y,\omega_z)=2\pi\times(2.7,92,74)$~Hz  located $350~\mu$m below the surface. It is formed by two currents on the chip and two currents in two pairs of Helmholtz coils. The currents are supplied by homebuilt current supplies with relative stability $<10^{-5}$ at 3~A \cite{reinhard2009design}. The final atom number is $N=2-4\times10^4$ and their temperature $T\sim80$~nK. The density is thus with $\bar{n}\approx 1.5\times 10^{11}$~atoms/cm$^3$ so low that the onset of Bose-Einstein condensation would occur at 5~nK. With $k_B T/\hbar \omega_{x,y,z}>20$ the ensemble can be treated by the Maxwell Boltzmann distribution. 
The trap lifetime $\gamma^{-1}=6.9$~s is limited by background gas collisions.
The clock transition is interrogated  via two-photon (microwave + radiofrequency) coupling, where the microwave  is detuned 500~kHz above the $\ket{1}$ to $\ket{F=2, m_F=0}$ transition (figure \ref{fig:clock_transition}). The microwave is coupled to a three-wire coplanar waveguide on the atom chip \cite{lacroute2010preliminary,bohi2010imaging}. The interaction of the atoms with the waveguide evanescent field allows to reach single photon Rabi frequencies of a few kHz with moderate power $\sim 0$~dBm. Since the microwave is not radiated, interference from reflections, that can lead to field-zeros and time varying phase at the atom position, is avoided. Thereby, the waveguide avoids the use of a bulky microwave cavity.
The microwave signal of fixed frequency $\nu_{MW}\sim6.8$~GHz is generated by a homebuilt synthesiser \cite{ramirez2010low} which multiplies a 100~MHz reference signal derived from an active hydrogen maser \footnote{The maser frequency is measured by the SYRTE primary standards. Its drift is at most a few $10^{-16}$ per day \cite{circularT} and can be neglected for our purposes.} to the microwave frequency without degradation of the maser phase noise. The actual phase noise is detailed in section \ref{sectionDickEffect}. The RF signal of variable frequency $\nu_{RF}\sim2$~MHz comes from a commercial DDS which supplies a "standard" wire parallel to the waveguide. The two-photon Rabi frequency is about $\Omega=3.2$~Hz making a $\pi/2$ pulse last $\tau_p=77.65~\text{ms}\gg 2\pi/\omega_z$. The pulse duration is chosen so that any Rabi frequency inhomogeneity, which was characterised in \cite{maineult2012spin}, is averaged out and Rabi oscillations show 99.5\% contrast.
Two pulses enclose a Ramsey time of $T_R=5$~s. 
Detection is performed via absorption imaging. A strongly saturating beam crosses the atom cloud and is imaged onto a back illuminated, high quantum efficiency CCD camera with frame transfer (Andor iKon M 934-BRDD). 
$20~\mu$s illumination without and with repump light, 5.5 ms and 8.5 ms after trap release, probes the $F=2$ and $F=1$ atoms independently. Between these two, a transverse laser beam blows away the $F=2$ atoms. Numerical frame re-composition  generates the respective reference images and largely reduces the effect of optical fringes \cite{ockeloen2010detection}. Calculation of the optical density and correction for the high saturation \cite{reinaudi2007strong} give access to the atom column density. The so found 2D atom distributions are fitted by Gaussians to extract the number of atoms in each state  $N_{1,2}$. The transition probability is calculated as $P=N_2/(N_1+N_2)$ accounting for total atom number fluctuations. The actual detection noise is discussed in section \ref{sectionDetection}. The total time of one experimental cycle is $T_c=16$~s.

\begin{center}
	\begin{tabular}{|c|c|}
	\hline
		Operation & Duration \\ \hline \hline
		MOT & 6.85 s \\  \hline
		compressed MOT & 20 ms \\ \hline
		optical molasses & 5 ms \\ \hline
		optical pumping & 1 ms \\ \hline
		magnetic trapping and compression & 230 ms \\ \hline
		RF evaporation & 3 s \\ \hline
		magnetic decompression & 700 ms \\ \hline
		first Ramsey pulse & 77.65 ms \\ \hline
		Ramsey time & 5 s \\ \hline
		second Ramsey pulse & 77.65 ms \\ \hline
		time of flight $(\ket{1}, \ket{2})$& (5.5 , 8.5) ms \\ \hline
		detection & $20~\mu$s\\ \hline
	\end{tabular}
\captionof{table}{Timing sequence of one experimental cycle. The total cycle time is 16~s.}
	\label{timingSequence}
\end{center}

\begin{figure}[h!]
\centering
\includegraphics[trim= 5mm 0 0 0, clip,width=1.00\columnwidth]{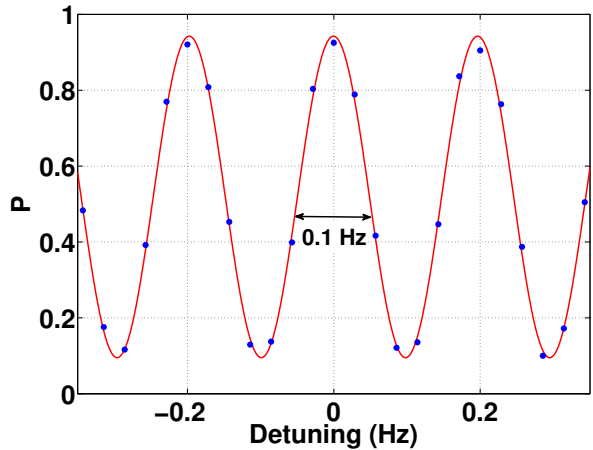}
\caption{Typical Ramsey fringes recorded at $T_R=5$~s while scanning the local oscillator detuning. Each point corresponds to a single experimental realization. One identifies the Fourier limited linewdith of 100~mHz and the very good contrast of 85\%.}.
\label{fig:RamseyFringes}
\end{figure}

\section{Stability measurement}
Prior to any stability measurement we record the typical Ramsey fringes. We repeat the experimental cycle while scanning $\nu_{LO}=\nu_{MW}+\nu_{RF}$ over $\sim3$ fringes. Doing so for various Ramsey times $T_R$ allows  to identify the central fringe corresponding to the atomic frequency $\nu_{at}$. Figure \ref{fig:RamseyFringes} shows typical fringes for $T_R=5$~s, where each point is a single shot. One recognises the Fourier limited linewidth of 100~mHz equivalent to $\sim10^{11}$ quality factor. The 85\% contrast is remarkable. A sinusoidal fit gives the slope at the fringe half-height $dP/d\nu=13.4/$Hz, which is used in the following stability evaluation to convert the detected transition probability into frequency.

Evaluation of the clock stability implies repeating the experimental cycle several thousand times. 
The clock is  free-running, i.e. we measure the transition probability at each cycle, but we do not feedback to the interrogation frequency $\nu_{LO}$. Only an alternation in successive shots from a small fixed negative to positive detuning, $\Delta_{mod}/(2\pi)=\pm 50$~mHz, probes the left and right half-height of the central fringe. The difference in $P$ between two shots gives the variation of the central frequency independent from long-term detection or microwave power drifts. In the longest run, we have repeated the frequency measurement over 18 hours.

\begin{figure}[h!]
\centering
\includegraphics[trim=0mm 0 00mm 0, clip,width=\columnwidth]{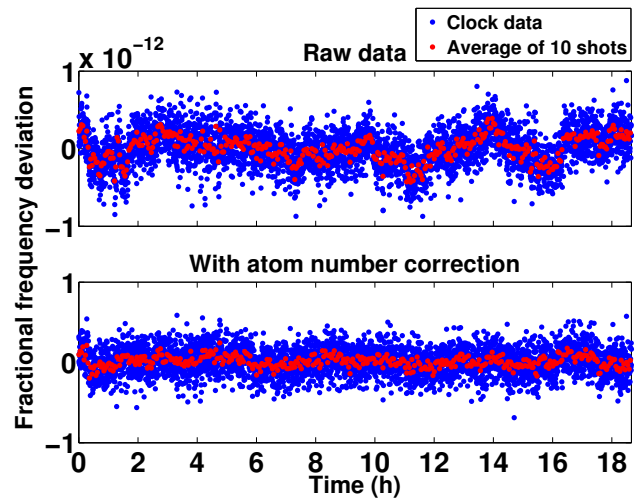}
\caption{(Color online) Relative frequency deviation when repeating the clock measurement over 18 h, (top) raw data, (bottom) after correction by the simultaneously detected total atom number. The blue dots represent single shots, red dots show an average of 10 shots.}
\label{fig:TACC_stability_shots}
\end{figure}

\begin{figure}[h!]
\centering
\includegraphics[trim=10mm 0 25mm 0,clip,width=\columnwidth]{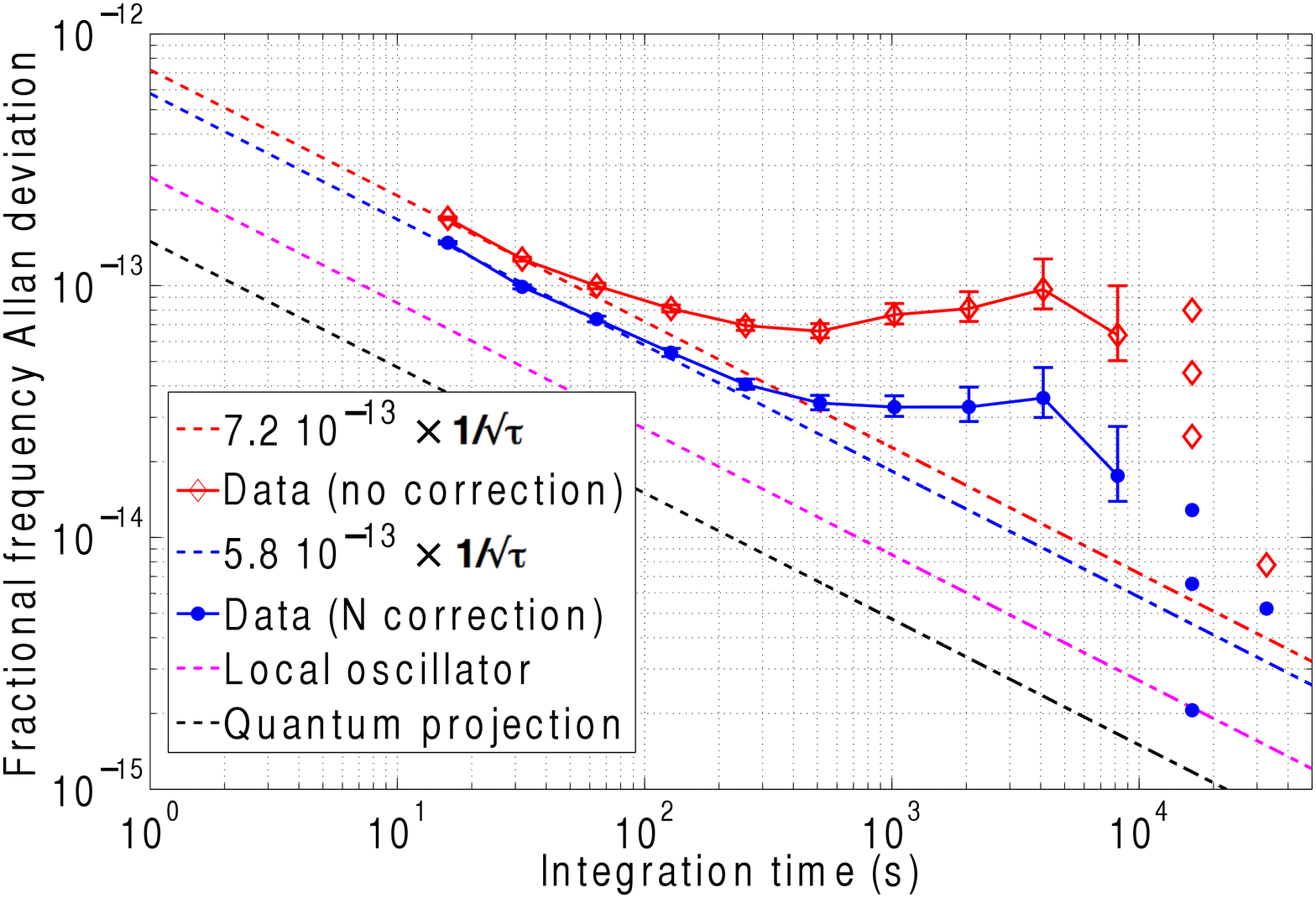}
\caption{(Color online) Allan standard deviation of the measured clock frequency with (blue circles) and without (red diamonds) atom number correction. For integration times smaller than $10^4$~s, the points and errorbars are calculated using the software "Stable32". Above $10^4$~s, the individual differences between successive packets of 1024 and 2048 measurements are given.
The $N$-corrected data follows initially $5.8\times10^{-13}\tau^{-1/2}$ (blue dashed line). The quantum projection noise and the local oscillator noise are given for reference.}.
\label{fig:TACC_stability_Allan}
\end{figure}

The measured frequency data is traced in figure \ref{fig:TACC_stability_shots} versus time. Besides shot-to-shot fluctuations one identifies significant long-term variations. Correction of the data with the atom number, by a procedure we will detail in section \ref{sectionNcorrection}, results in substantial improvement. 
We analyze the data by the Allan standard deviation which is defined as \cite{allan1966statistics}
\begin{equation}
\sigma_y^2(\tau)\equiv\frac{1}{2}\sum_{k=1}^{\lfloor L/2^l\rfloor-1}\left(\bar{y}_{k+1}-\bar{y}_k\right)^2.
\label{AllanVariance}
\end{equation}
Here $L$ is the total number of data points and the $\bar{y}_k$ are averages over packets of $2^l$ successive data points with $l\in\{0,1,...\lfloor\log_2L\rfloor\}$ and $\tau=2^l T_c$. 
Figure \ref{fig:TACC_stability_Allan} shows the Allan standard deviation of the uncorrected and corrected data. For $0\leq l\leq9$ the points and their errorbars are plotted as calculated with the software "Stable32" \footnote{http://www.wriley.com}. This software uses equation \ref{AllanVariance} to find the points. The error bars are calculated as the 5\% - 95\% confidence interval based on the appropriate $\chi^2$ distribution. The software stops  output at $l=\lfloor\log_2L\rfloor-2$ since there are too few differences $\bar{y}_{k+1}-\bar{y}_{k}$ to give a statistical errorbar. Instead we directly plot  all differences for $l=10$ and 11.

The Allan standard deviation shows the significant improvement brought by the atom number correction. The uncorrected data starts at $\tau=T_c=16$~s with $\sigma_y=1.9\times10^{-13}$ shot-to-shot. For the $N$-corrected data, the shot-to-shot stability is $\sigma_y=1.5\times10^{-13}$. Up to $\tau\approx100$~s the corrected frequency fluctuations follow a white noise behaviour of $\sigma_y(\tau)=5.8\times 10^{-13}\tau^{-1/2}$. 
At $\tau\approx1000$~s, the fluctuations are above the $\tau^{-1/2}$ behaviour but decrease again at $\tau>5000$~s. For $\tau>10^4$~s, 3 of the 4 individual differences are below $10^{-14}$. This lets us expect that a longer stability evaluation would indeed confirm a stability in the $10^{-15}$ range with sufficient statistical significance. 
The "shoulder" above the white noise behaviour is characteristic for an oscillation  at a few $10^3$~s half-period. Indeed, this oscillation is visible in the raw data in figure \ref{fig:TACC_stability_shots}. Its cause is yet to be identified through simultaneous tracking of many experimental parameters - a task which goes beyond the scope of this paper.

Table \ref{noise_budget} gives a list of identified shot-to-shot fluctuations that contribute to the clock frequency noise. Treating them as statistically independent and summing their squares gives a fractional frequency fluctuation of $1.5\times 10^{-13}$ shot-to-shot or $6.0\times 10^{-13}$ at 1~s, corresponding to the measured stability. We have thus identified all major noise sources building a solid basis for future improvements. In the following we  discuss each noise contribution in detail.

\begin{center}
	\begin{tabular}{|c|c|c|}
	\hline
		Relative frequency stability ($10^{-13}$) & shot-to-shot & at 1~s \\ \hline \hline
		measured, without correction & $2.0$ & $7.2$ \\  \hline
		\bf{ measured, after $N$ correction} & $\mathbf{1.5}$& $\mathbf{5.8}$ \\  \hline \hline
		atom temperature  & $1.0$ & $3.9$ \\ \hline
		magnetic field  & $0.7$ & $2.6$ \\ \hline
		local oscillator & $0.7$ & $2.7$ \\ \hline
		quantum projection & $0.4$ &$1.5$ \\  \hline
	  $N$ correction & $0.4$ & $1.5$ \\ \hline
		atom loss & $0.3$ & $1.1$ \\ \hline
		detection & $0.3$ & $1.1$ \\ \hline \hline
		{total estimate} &  ${1.5}$ & ${6.0}$ \\ \hline
	\end{tabular}
		\captionof{table}{List of identified contributions to the clock (in)stability. Atom temperature fluctuations dominate followed by magnetic field fluctuations and local oscillator noise. The quadratic sum of all contributions explains the measured stability.}
	\label{noise_budget}
\end{center}

\section{Noise Analysis}

In a passive atomic clock, an electromagnetic signal generated by an external local oscillator (LO) interacts with an atomic transition. 
 The atomic transition frequency $\nu_{at}$ is probed by means of spectroscopy. 
The detected atomic excitation probability $P$ is either used to correct the LO on-line such that $\nu_{LO}=\nu_{at}$, or, as applied here, the LO is left free-running and the measured differences $(\nu_{LO}-\nu_{at})(t)$ are recorded for post-treatment. The so calibrated LO signal is the useful clock output. 

When concerned with the stability of the output frequency, we have to analyze the noise of each element within this feed-back loop, i.e.

A. noise from imperfect detection,

B. folded-in fluctuations of the LO frequency known as Dick effect,

C. fluctuations of the atomic transition frequency induced by interactions with the environment or between the  atoms. 

We begin by describing the most intuitive contribution (A. detection noise) and finish by the most subtle (C. fluctuations of the atomic frequency).

\subsection{Detection and quantum projection noise}
\label{sectionDetection}
The clock frequency is deduced from absorption imaging the atoms in each clock state as described in section \ref{sectionSetup}.  $N_1$ and $N_2$ are obtained by fitting Gaussians to the atom distribution, considering a square region-of-interest of $\sim3\times3$  cloud widths.  

Photon shot noise and optical fringes may lead to atom number fluctuations of standard deviation $\sigma_{det}$. These fluctuations add to the true atom number.  Analyzing blank images, we confirm that $\sigma_{det}^2$ increases as the number of pixels in the region-of-interest and that optical fringes have been efficiently suppressed \cite{ockeloen2010detection}. This scaling has led to the choice of short times-of-flight where the atoms occupy fewer pixels \footnote{The minimum time-of-flight is given by the onset of optical diffraction at high optical density.}.   Supposing the same  $\sigma_{det}$ for both states, we find for the transition probability noise $\sigma_{P,det}=\sigma_{det}/(\sqrt{2}N)$ with $N=N_1+N_2$.

Another $P$ degradation  may occur if the Rabi frequency of the first pulse fluctuates  or if the detection efficiency varies between the $\ket{1}$ and $\ket{2}$ detection. The latter  may arise from  fluctuations of the detection laser frequency on the timescale of the 3~ms difference in time-of-flight. Both fluctuations induce a direct error $\sigma_{P,Rf+lf}$ on $P$ independent from the atom number. 

Quantum projection noise is a third cause for fluctuations in $P$. This fundamental noise arises from the fact that the detection projects the atomic superposition state onto the base states. Before detection, the atom is in a near-equal superposition of $\ket{1}$ and $\ket{2}$. The projection then can result in either base state with equal probability giving $\sigma_{QPN}=1/2$ for one atom. Running the clock with $N$ (non-entangled) atoms  is equivalent to $N$ successive measurement resulting in $\sigma_{P,QPN}=1/(2\sqrt{N})$ shot-to-shot. 

\begin{figure}[h!]
\centering
\includegraphics[width=90mm]{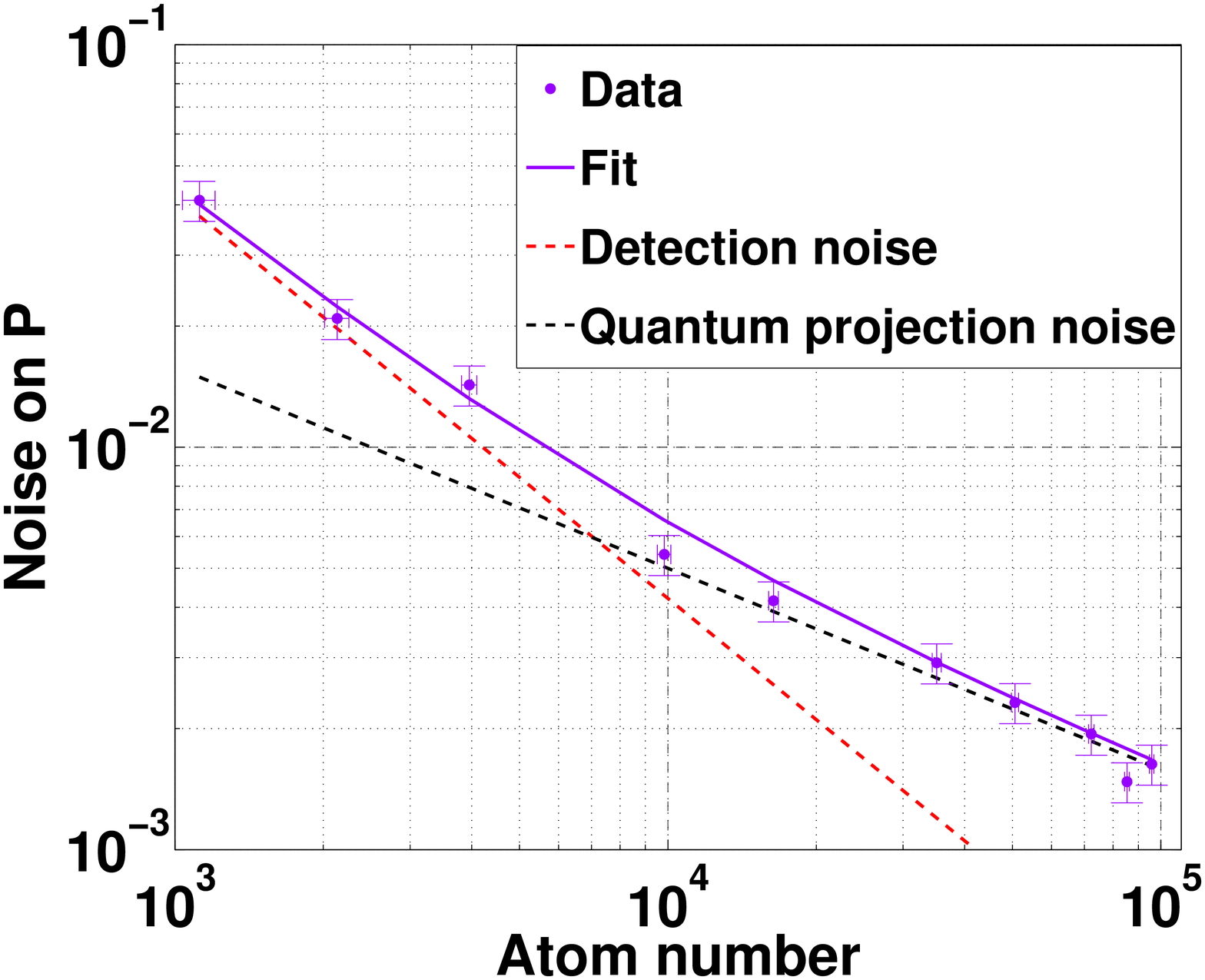}
\caption{Characterization of the detection noise. Only a single $\pi/2$ pulse is applied and $P=N_2/N$ detected. The shot-to-shot Allan deviation  is plotted as a function of the total atom number. We fit the data with the quadratic sum of the detection noise $\sigma_{det}/(\sqrt{2}N)$, the quantum
projection noise $1/(2\sqrt{N})$ and the Rabi frequency and laser frequency noise $\sigma_{P,Rf+lf}$. The fit gives $\sigma_{det}=59$ atoms and $\sigma_{P,Rf+lf}<10^{-4}$.}
\label{fig:detprep}
\end{figure}

We quantify the above three noise types from an independent measurement: Only the first $\pi/2$ pulse is applied and $P$ is immediately detected. The measurement is repeated for various atom numbers and $\sigma_P(N)$ is extracted. Figure \ref{fig:detprep} shows the measured $\sigma_P$ shot-to-shot versus $N$. Considering the noise sources as statistically independent, we fit the data by $\sigma_P^2=\sigma_{det}^2/(2N^2)+1/(4N)+\sigma_{P, RF+lf}^2$ and find $\sigma_{det}=59$~atoms  and $\sigma_{P, RF+lf}<10^{-4}$.  $\sigma_{det}$ is equivalent to an average of $2.2$~atoms/pixel for our very typical absorption imaging system. The low $\sigma_{RF+lf}$ proves an excellent passive microwave power stability  $<2.5\times10^{-4}$, which may be of use in other experiments, in particular  microwave dressing \cite{bohi2009coherent,sarkany2014controlling}.

During the stability measurement of figure \ref{fig:TACC_stability_shots} about $20\,000$ atoms are detected, which is equivalent to $\sigma_{y,QPN}=0.4\times10^{-13}$ shot-to-shot. The detection region-of-interest is slightly bigger than for the above characterisation, so that $\sigma_{det}=69$~atoms, corresponding to $\sigma_{y,det}=\sigma_{det}(\nu_{at}N)^{-1} |dP/d\nu|^{-1}=\frac{\sigma_{det}}{\sqrt{2} N}\frac{1}{\nu_{at} 13.4}=0.3\times10^{-13}$ shot-to-shot. 
In both we have used  $dP/d\nu$ as measured in figure \ref{fig:RamseyFringes}.

\subsection{Local oscillator noise}
\label{sectionDickEffect}
The experimental cycle probes $\nu_{at}-\nu_{LO}$ only during the Ramsey time. Atom preparation and detection cause dead time. Repeating the experimental cycle then constitutes periodic sampling of the LO frequency and its fluctuations. This, as well-known from numerical data acquisition, leads to aliasing. It folds high Fourier frequency LO noise close to multiples of the sampling frequency $1/T_c$ back to low frequency variations, which  degrade the clock stability. Thus even high Fourier frequency noise can degrade the clock signal. The degradation is all the more important as the dead time is long and the duty cycle $d=T_R/T_c$ is low.
This stability degradation $\sigma_{y,Dick}$ is known as the Dick effect \cite{dick1987local}. It
 is best calculated using the sensitivity function $g(t)$ \cite{santarelli1998frequency}:  during dead-time, $g=0$ whereas during $T_R$, when the atomic coherence $\ket{\psi}=(\ket{1}+e^{i\phi}\ket{2})/\sqrt{2}$ is fully established $g=1$. During the first Ramsey pulse, when the coherence builds up, $g$ increases as $\sin\Omega t$ for a square pulse  and decreases symmetrically for the second pulse \footnote{The sensitivity function can be understood by visualising the trajectory of a spin 1/2 on the Bloch sphere.}. Then the interrogation outcome is
\begin{equation}
 \delta\nu=\frac{\int^{T_c/2}_{-T_c/2}\left(\nu_{at}(t)-\nu_{LO}(t)\right)g(t)\,dt}{\int^{T_c/2}_{-T_c/2}g(t)\,dt}
 \end{equation}
  with
\begin{equation}
	g(t)=
		\begin{cases}
			a \sin{\Omega (T_R/2 +  \tau_p +t)} & -\tau_p-\frac{T_R}{2}\le t \le -\frac{T_R}{2} \\
			a \sin{\Omega \tau_p} & -\frac{T_R}{2} \le t < \frac{T_R}{2}  \\
			a \sin{\Omega (\frac{T_R}{2} +  \tau_p -t)} & \frac{T_R}{2}  \le t \le \frac{T_R}{2} +  \tau_p \\
			0 & \text{otherwise}.
		\end{cases}
\end{equation}
Typically $\Omega\tau_p=\pi/2$ and, for operation at the fringe half-height, $a=\sin{\Delta_{mod} T_R}=1$. Because of the periodicity of repeated clock measurements, it is convenient to work in Fourier space with
\begin{equation}
		 g_l=\frac{1}{T_c} \int_{-T_c/2}^{T_c/2} g(t)  \cos(2 \pi l\, t/T_c) dt. 
\end{equation}
Using  the power spectral density of the LO frequency noise $S_y^f(f)$, the contribution to the clock stability  becomes the quadratic sum over all harmonics  \cite{santarelli1998frequency}
\begin{equation}
\sigma^2_{y,Dick}(\tau)=\frac{1}{\tau}\sum_{l=1}^{\infty}\left(\frac{{g_l}}{g_0}\right)^2 S_y^f(l/T_C).
\label{eqnDickEffect}
\end{equation}
Here we have assumed $\nu_{at}$ constant in time; its fluctuations are treated in the next section. The coefficients $(g_l/g_0)^2$ are shown as points in figure \ref{fig:PSDMaser} for our conditions. The weight of the first few harmonics is clearly the strongest, rapidly decaying over 6 decades in the range $1/T_c\approx0.1$~Hz to $1/\tau_p\approx 10$~Hz. Above $\sim10$~Hz the $g_l$ are negligible.

To measure $S_y^f(f)$ we divide our LO into two principal components: the 100~MHz reference signal derived from the hydrogen maser and the frequency multiplication chain generating the $6.8$~GHz interrogation signal. We characterise each independently by measuring the phase noise spectrum $S_{\phi}(f)$. The fractional frequency noise $S_y^f(f)$  is obtain from  simple differentiation as $S_y^f(f)=f^2S_\phi(f)/\nu_{MW}^2 $ \cite{santarelli1998frequency}. The frequency noise of the  RF signal can be neglected as its relative contribution is 3 orders of magnitude smaller.

\begin{figure}[h!]
\centering
\includegraphics[width=90mm]{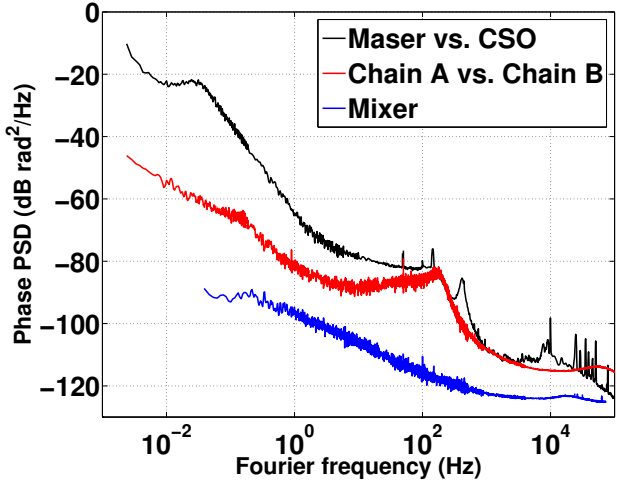}
\caption{(Color online) Phase noise power spectral density of the local oscillator. The frequency multiplication chain and the 100 MHz reference signal are characterised separately. The beat between two quasi identical chains is performed at 6.8~GHz (red). The beat of the reference signal against a cryogenic sapphire oscillator is taken at 100~MHz and scaled to 6.8~GHz (black). The noise of the reference signal dominates in the low frequency part, where our clock is sensitive. Both results are above the intrinsic noise of the measurement system (blue).}
\label{fig:PSD_synthesizer_maser}
\end{figure}

\begin{figure}[h!]
\centering
\includegraphics[width=\columnwidth]{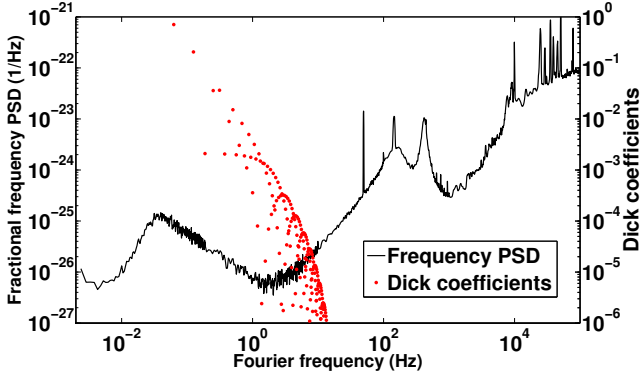}
\caption{(Color online) (black) Same data as figure \ref{fig:PSD_synthesizer_maser} now expressed as fractional frequency fluctuations $S_y=f^2 S_{\phi}/ \nu_{MW}^{2}$. (red)  Fourier coefficients of the sensitivity function $(g_l/g_0)^2$ for our conditions ($T_c=16$~s, $T_R=5$~s and $\tau_p=77.65$~ms). Multiplication of the two gives the stability degradation known as Dick effect.}
\label{fig:PSDMaser}
\end{figure}

We characterize the frequency multiplication chain by comparing it to a second similar model also constructed in-house.
The two chains are locked to a common 100~MHz reference and their phase difference at 6.834~GHz is measured as DC signal using a phase detector (Miteq DB0218LW2) and a  FFT spectrum analyzer (SRS760). The measured $S_{\phi}(f)$ is divided by 2 assuming equal noise contributions from the two chains. 
It is shown in figure \ref{fig:PSD_synthesizer_maser}.  It features a $1/f$ behaviour up to $f=10$~Hz and reaches a phase flicker floor of $-115$~dB~rad$^2$/Hz at 1~kHz. The peak at $f=200$~Hz is due to the phase lock of a 100~MHz  quartz inside the chain  to the reference signal. As we will see in the following, its contribution to the Dick effect is negligible.

The 100~MHz reference signal is generated by a 100~MHz quartz locked to a 5~MHz quartz locked with  40~mHz bandwidth  to an active hydrogen maser (VCH-1003M). We measure this reference signal   against a 100~MHz signal derived from a cryogenic sapphire oscillator (CSO) \cite{luiten1995power,mann2001cryogenic}. Now the mixer is M/A-COM PD-121. The CSO is itself locked to the reference signal but with a time constant of $\sim 1000$~s \cite{guena2012progress}. This being much longer than our cycle time, we can, for our purposes, consider the two as free running.
The CSO is known from prior analysis \cite{chambon2005design} to be at least 10~dB lower in phase noise than the reference signal for Fourier frequencies higher than 0.1~Hz. Thus the measured noise can be attributed to the reference signal for the region of the spectrum $f>1/T_c$ where our clock is sensitive.  
The phase noise spectrum is shown in figure  \ref{fig:PSD_synthesizer_maser}. For comparison it was scaled to 6.8 GHz by adding 37~dB. Several maxima characteristic of the several phase locks in the systems can be identified.  At low Fourier frequencies, the reference signal noise is clearly above the chain noise. For all frequencies, both are well above the noise floor of our measurement system. The noise of the reference signal being dominant in the range $1/T_c$ to $1/\tau_p$, where our clock is sensitive, we neglect the chain noise in the following.

Using equation \ref{eqnDickEffect}, we estimate the Dick effect contribution as  $\sigma_{y,Dick}=2.7 \times 10^{-13} \tau^{-1/2}$. 
This represents the second biggest contribution to the noise budget (table \ref{noise_budget}). It is due to the important dead time and the long cycle time which folds-in the LO noise spectrum where it is strongest. Improvement is possible, first of all, through  reduction of the dead time which is currently dominated by the $\sim7$~s MOT loading phase and the 3~s evaporative cooling. Options for faster loading include pre-cooling in a 2D MOT \cite{dieckmann1998two} or a single-cell fast pressure modulation  \cite{dugrain2014alkali}. Utilization of a better local oscillator like the cryogenic sapphire oscillator seems obvious but defies the compact design. Alternatively, generation of low phase noise microwaves from an ultra-stable laser and femtosecond comb has been demonstrated by several groups \cite{bartels2005femtosecond,millo2009ultra, kim2010microwave} and on-going projects aim at miniaturisation of such systems \cite{del2007optical}. If a quartz local oscillator remains the preferred choice, possibly motivated by cost, one long Ramsey time must be divided into several short interrogation intervals interlaced by non-destructive detection \cite{westergaard2010minimizing,bernon2011heterodyne,lodewyck2009nondestructive}.


\begin{figure}[h!]
\centering
\includegraphics[width=\columnwidth]{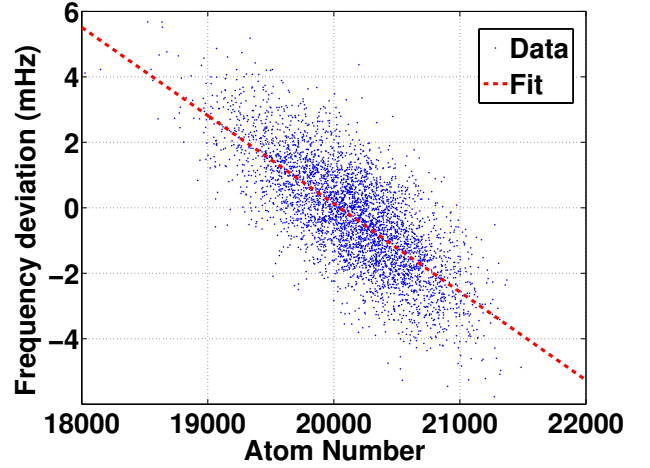}
\caption{Correlation between the detected atom number and the clock frequency for the data of figure \ref{fig:TACC_stability_shots}. Fitting with a linear regression gives $k=-2.70(7)~\mu$Hz/atom, which allows to correct the clock frequency  at each shot and yields substantial stability improvement.}
\label{fig:atom_freq_correlation}
\end{figure}

\subsection{Fluctuations of the atomic frequency}
\subsubsection{Atom number fluctuation}
\label{sectionNcorrection}

Having characterised the fluctuations of the LO frequency, we now turn to fluctuations of the atomic frequency. We begin by atom number fluctuations. Due to the  trap confinement and the ultra-cold temperature, the atom density is 4 orders of magnitude higher than what is typically found in a fountain clock. Thus the effect of atom-atom interactions  on the atomic frequency must be taken into account even though $^{87}$Rb presents a substantially lower collisional shift than the standard $^{133}$Cs. Indeed, when plotting the measured clock frequency against the detected atom number $N=N_1+N_2$, which fluctuates by 2-3\% shot-to-shot, we find a strong correlation (figure \ref{fig:atom_freq_correlation}). The  distribution is compatible with a linear fit with slope  $k=-2.70(7)~\mu$Hz/atom.
In order to confirm this value with a theoretical estimate we use the mean field approach  and the s-wave scattering lengths $a_{ij}$ which depend on the atomic states only \cite{harber2002effect}
\begin{equation}
\begin{split}
\Delta\nu_C(\vec{r})&=\frac{2 \hbar}{m} n(\vec{r}) \left[(a_{22}-a_{11})+(2 a_{12} - a_{11} - a_{22}) \theta\right].\\
\end{split}
\label{collisional_shift}
\end{equation}
$n(\vec{r})$ is the position dependent density and $a_{11}=100.44 a_0$, $a_{22}=95.47 a_0$, $a_{12}=98.09 a_0$ are the scattering lengths with $a_0=0.529\times 10^{-10}$~ m \cite{harber2002effect}. We assume perfect $\pi/2$ pulses and so 
$\theta\equiv(N_1-N_2)/N=0$.
Integrating over the Maxwell-Boltzmann density distribution  we get
\begin{equation}
\overline{\Delta\nu_C}=N \times \frac{-\hbar (a_{11}-a_{22}) \sqrt{m} \omega_x \omega_y \omega_z}{4 (\pi k_B T)^{3/2}}.
\label{collisional_shift}
\end{equation}
We must  consider that the atom number decays during the $T_R=5$~s since the trap life time is $\gamma^{-1}=6.9$~s. We replace $N$ by its temporal average
\begin{equation}
\begin{split}
\overline{N}&=\frac{1}{T_R} \int_0^{T_R} N_i e^{-\gamma t}  dt \\
&= N_i\frac{1-e^{-\gamma T_R}}{\gamma T_R} \\
&= N_f\frac{e^{\gamma T_R}-1}{\gamma T_R} \\
&\approx 1.47\; N_f
\end{split}
\label{collisionalShift}
\end{equation}
where $N_i$ and $N_f$ are the initial and final atom numbers, respectively. Note that $N_f$ is the  detected atom number.
Using $T=80$~nK, which is compatible with an independent measurement, we recover the experimental collisional shift of $k=-2.7~\mu$Hz/(detected atom). It is equivalent to an overall collisional shift of $\overline{\Delta\nu_C}=-54$~mHz for $N_f=20\,000$.

Using $k$ and the number of atoms detected at each shot we can correct the clock frequency for fluctuations. The corrected frequency is given in figure \ref{fig:TACC_stability_shots}  showing a noticeable improvement in the short-term and long-term stability. The Allan deviation indicates a clock stability of $5.8	\times10^{-13}\tau^{-1/2}$ at short term  as compared to $7.2	\times10^{-13}\tau^{-1/2}$ for the uncorrected data. At long term the improvement is even more pronounced.
This demonstrates the efficiency of the $N$-correction. Furthermore, the experimentally found $k$ shows perfect agreement with our theoretical prediction so that the theoretical coefficient can in future be used from the first shot on without the need for post-treatment.

While we have demonstrated the efficiency of the atom number correction, the procedure has imperfections for two reasons: The first, of technical origin, are fluctuations in the atom number detectivity as evaluated in section \ref{sectionDetection}. The second arises from the fact that atom loss from the trap is a statistical process. For the first, we get $\sigma_{y,correction}=\sqrt{2}|k| \sigma_{det}/\nu_{at}=0.4\times10^{-13} $ shot-to-shot.  
This value is well below the measured clock stability, but may become important when other noise sources are eliminated. It can be improved by reducing the atom density and thus $k$ or by better detection, in particular at shorter time-of-flight where the camera region-of-interest can be smaller.
The second cause, the statistical nature of atom loss, translates into fluctuations that in principle cannot be corrected. The final atom number $N_f$ at the end of the Ramsey time is known from the detection, but the initial atom number $N_i$ can only be retraced with a statistical error.
To estimate this contribution we first consider the decay from the initial atom number $N_i$. At  time $t$, the probability for a given atom to still be trapped is $e^{-\gamma t}$ and the
probability to have left the trap is $1-e^{-\gamma t}$. Given $N_i$, the probability $p$ to have $N_t$ atoms at $t$ is proportional
to $e^{-N_t \gamma t} \left(1-e^{-\gamma t}\right)^{N_i-N_t}$ and to the number of possible combinations:
\begin{equation}
	p(N_t \text{ given } N_i)=\frac{N_i!}{N_t! (N_i-N_t)!}e^{-N_t \gamma t} (1-e^{-\gamma t})^{N_i-N_t}.
	\label{symmetric_losses}
\end{equation}
The sum of this binomial distribution over all $0 \leq N_t \leq N_i$ is by definition normalised. We are interested in the opposite case: since we detect $N_f$ at $t=T_R$, we search the probability of  $N_t$ given $N_f$. 
\begin{equation}
\begin{split}
	p(N_t \text{ given } N_f)&= \frac{A N_t!}{N_f! (N_t-N_f)!}\\
	&\times e^{-N_f \gamma (T_R-t)} (1-e^{-\gamma (T_R-t)})^{N_t-N_f}.
\end{split}
	\label{symmetric_losses_distribution}
\end{equation}
The combinatorics are as in equation \ref{symmetric_losses} when replacing $N_t\rightarrow N_f$ and $N_i\rightarrow N_t$, but now normalisation sums over $0 \leq N_t < \infty$. 
Here it is convenient to approximate the binomial distribution  by the normal distribution 
\begin{equation}
	p(N_t \text{ given } N_f)\approx \frac{A}{\sqrt{2\pi \eta}}e^{-(N_f-N_te^{-\gamma (T_R-t)})^2/(2\eta)}
\end{equation}
with $\eta=N_t e^{-\gamma (T_R-t)} (1-e^{-\gamma (T_R-t)})$ and hence  $A=e^{-\gamma (T_R-t)}$.
Then, the  mean of $N_t$ is 
\begin{equation}
\left<N_t\right>=(N_f+1) e^{\gamma (T_R-t)}-1\approx N_f e^{\gamma(T_R-t)}\\
\end{equation}
and its statistical error 
\begin{equation}
\begin{split}
\sigma_{N_t}&=\sqrt{(1-e^{\gamma(T_R-t)})(2-(N_f+2) e^{\gamma(T_R-t)})}\\
&\approx\sqrt{N_f (e^{\gamma(T_R-t)}-1)e^{\gamma(T_R-t)}}.
\end{split}
\end{equation}
Setting $t=0$, we get $\sigma_{N_i}=210$.
Integrating $\sigma_{N_t}$ over $T_R$ gives $\overline{\sigma_{N}}=113\approx \sigma_{N_i}/2$ and a frequency fluctuation of $\sigma_{y,losses}=0.3\times10^{-13}$ shot-to-shot.
This can be improved by increasing the trap lifetime well beyond the Ramsey time, which for our set-up  implies better vacuum with lower background pressure. Alternatively one can perform a non-destructive measurement of the initial atom number \cite{kohnen2011minimally}. 
Assuming an error of 80 atoms on such a detection  would decrease the frequency noise to $\sigma_{y,losses}=0.1\times10^{-13}$ shot-to-shot.

\subsubsection{Magnetic field and atom temperature fluctuations}

We have analyzed the effect of atom number fluctuations. Two other parameters strongly affect the atomic frequency: the atom temperature and the magnetic field. We show that their influence can be evaluated by measuring the clock stability for different magnetic fields at the trap center. We begin by modelling the dependence of the clock frequency.

\begin{figure}[h!]
\centering
\includegraphics[trim=0 0 20mm 0, clip, width=\columnwidth]{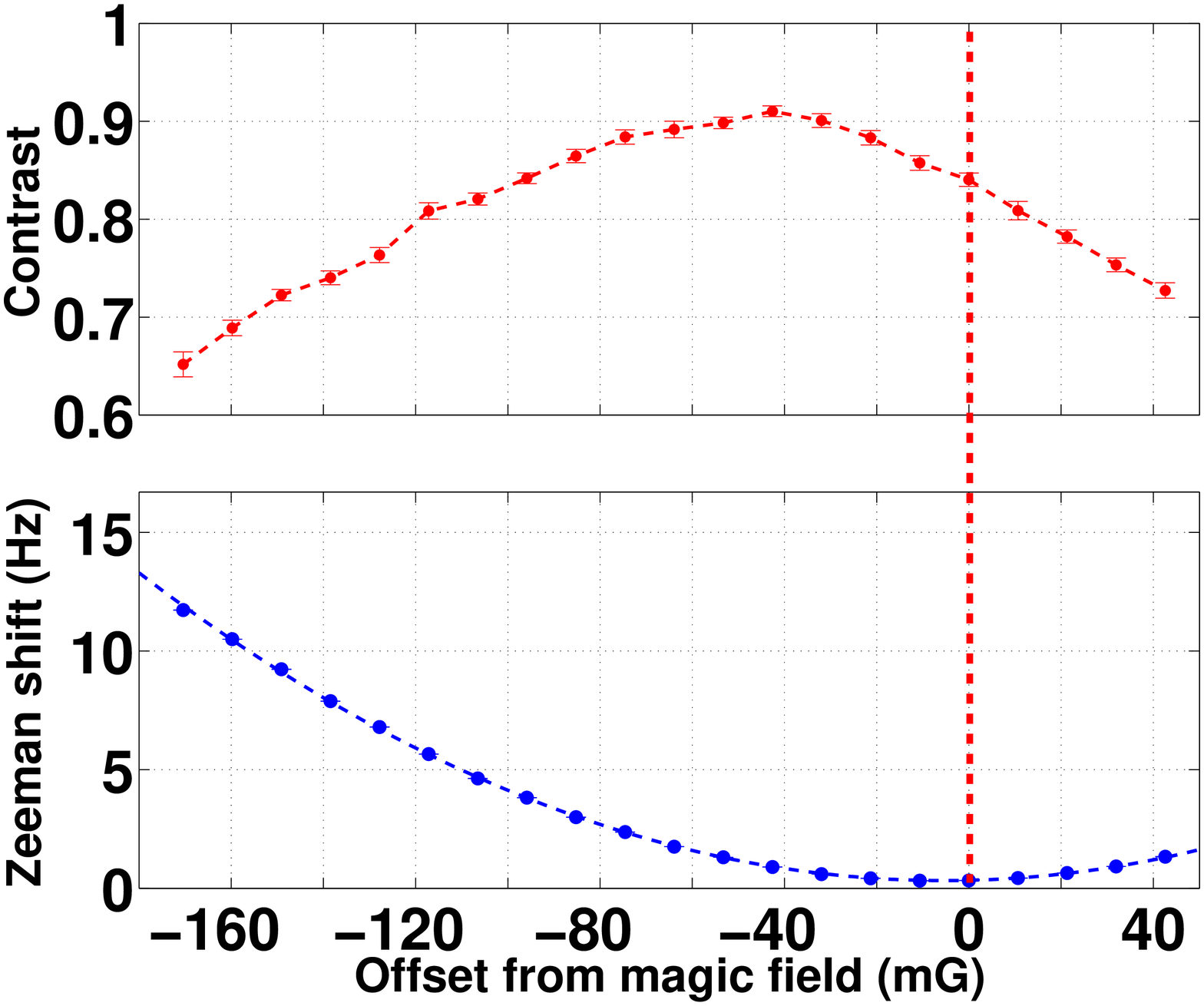}
\caption{Fringe contrast (top) and differential Zeeman shift (bottom) of the clock frequency with respect to the frequency minimum for various magnetic fields. $\delta B=0$ indicates the magic field of 3.229~G. The contrast maximum is offset by $-40~\text{mG}$.}
\label{fig:contrast_frequency}
\end{figure}

Our clock operates near the magic field $B_m\approx3.229$~G for which the transition frequency has a minimum of -4497.31~Hz  with respect to the field free transition,
\begin{equation}
\Delta\nu_B(\vec{r})=b(B(\vec{r})-B_m)^2
\end{equation}
with $b\approx 431~\text{Hz/G}^2$.
 For atoms trapped in a harmonic potential in the presence of gravity, the Zeeman shift becomes position dependent 
\begin{equation}
\Delta\nu_B(\vec{r})=\frac{b m^2}{\mu_B^2}\left(\omega_x^2 x^2 + \omega_y^2 y^2 + \omega_z^2 z^2 -2gz +\delta B \frac{\mu_B}{m}\right)^2
\end{equation}
with $\delta B\equiv B(\vec{r}=0)-B_{m}$ and $g$ the gravitational acceleration \cite{rosenbusch2009magnetically}. 
Using the Maxwell-Boltzmann distribution the ensemble averaged Zeeman shift is
\begin{equation}
\begin{split}
\overline{\Delta \nu_B} =\frac{b}{\mu_B^2}(&\frac{4 g^2 m k_B T}{\omega_z^2}+15 k_B^2 T^2  \\
&+ 6 \mu_B \delta B k_B T +\delta B^2 \mu_B^2  ).\\
\end{split}
\label{averageZeemanShift}
\end{equation}
Differentiation with respect to $\delta B$ leads to the effective magic field 
\begin{equation}
\delta B^B_{0}=\frac{-3 k_B T}{\mu_B}
\end{equation}
where the ensemble averaged frequency is  independent from magnetic field fluctuations.
For $T=80$~nK, $\delta B^B_0=-3.6$~mG whose absolute value almost coincides with the  magnetic field inhomogeneity across the cloud $\left(\overline{B(\vec{r})^2-\overline{B}^2}\right)^{1/2}= \sqrt{6} k_B T/\mu_B=2.92$~mG.  $\delta B^B_0$ is close to 
the field of maximum contrast $\delta B_0^C\approx-40$~mG  such that the fringe contrast is still 85\% (figure \ref{fig:contrast_frequency}).

If $\delta B\neq\delta B_0^B$ is chosen the clock frequency fluctuations due to magnetic field fluctuations are
\begin{equation}
\sigma_{y,B}=\frac{2 b}{\nu_{at}} \left|\delta B_0^B-\delta B\right| \sigma_B.
\label{sigmaB}
\end{equation}
We will use this dependence to measure $\sigma_B$.

Temperature fluctuations affect the range of magnetic fields probed by the atoms  and the atom density, i.e. the collisional shift.
Differentiation of both with respect to temperature also leads to an extremum, where the clock frequency is insensitive to temperature fluctuations. The extremum puts a concurrent condition on the magnetic field with
\begin{equation}
\begin{split}
\delta B^T_{0}=&-\frac{15 k_B T+\frac{2 g^2 m}{\omega_z^2}}{3 \mu_B}\\
&-\frac{\hbar (a_{11}-a_{22}) (e^{\gamma T_R } -1) \sqrt{m}N_f \mu_B \omega_x \omega_y \omega_z}{16  \pi^{3/2} b (k_B T)^{5/2}  \gamma T_R }.
 \end{split}
\end{equation}
For our conditions, $\delta B_0^B=-3.6$~mG and $\delta B_0^T=-79$~mG are not identical but close and centered around $\delta B_0^C$. We will see in the following that a compromise can be found where the combined effect of magnetic field and temperature fluctuations is minimised. A "doubly magic" field can not be found as always $\delta B_0^T<\delta B_0^B$, but lower $T$ reduces their difference.
If $\delta B\neq\delta B_0^T$ is chosen, the clock frequency fluctuations due to temperature fluctuations are
\begin{equation}
\sigma_{y,T}=\frac{6 b k_B}{\mu_B\nu_{at}} \left| \delta B_0^T-\delta B\right|\sigma_T
\label{sigmaT}
\end{equation}
thus varying $\delta B$ allows to measure $\sigma_T$, too.

\begin{figure}[h!]
\centering
\includegraphics[width=95mm]{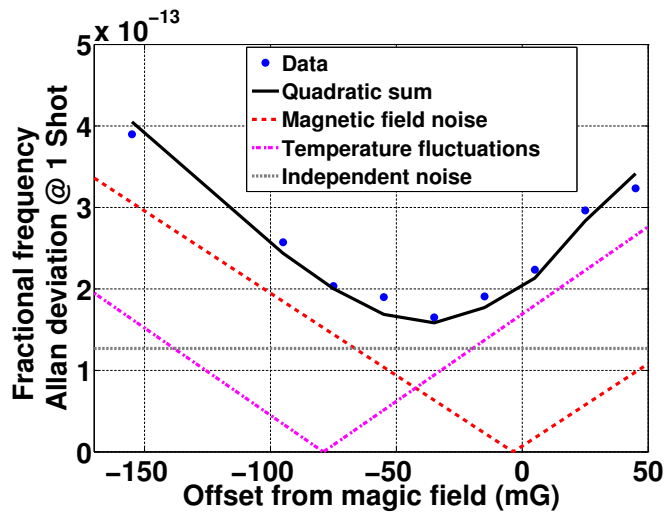}
\caption{(Color online) Shot-to-shot clock stability for various magnetic fields. Error bars are smaller than the point size.
One observes a clear optimum at $\delta B=-35$~ mG. Fitting with the quadratic sum of all identified noise contributions allows to quantify the atom temperature fluctuations (0.4~nK shot-to-shot) and magnetic field fluctuations ($16~\mu$G shot-to-shot). The individual contributions are shown as dashed lines. Two sweet spots exist where the temperature dependence and the magnetic field dependence vanish.}
\label{fig:BTnoisefit}
\end{figure}

We determine $\sigma_B$ and $\sigma_T$ experimentally by repeating several stability measurements for different $\delta B$ over a range of 200 mG  where the contrast is
above 70\%. The shot-to-shot stability  is shown in figure \ref{fig:BTnoisefit}.
One identifies a clear minimum of the instability at $\delta B\approx-35$~mG, which coincides with $\delta B_0^C$ and is a compromise between the two optimal points $\delta B_0^T$ and $\delta B_0^B$. This means, that both magnetic field and temperature fluctuations are present with roughly equal weight. We model the data with a quadratic sum of all so far discussed noise sources. Most of them give a constant offset; the slight variation due to the contrast variation shown in figure \ref{fig:contrast_frequency} is negligible. $\sigma_{y,B}$ and $\sigma_{y,T}$ are fitted by adjusting $\sigma_B$ and $\sigma_T$. We find shot-to-shot  temperature fluctuations of $\sigma_T=0.44$~nK or 0.55\% relative to 80~nK. The shot-to-shot  magnetic field fluctuations  are $\sigma_B=16~\mu$G or $5\times 10^{-6}$ in relative units. The values demonstrate our exceptional control of the experimental apparatus. Because the ambient magnetic field varies by $< 10$~mG and the lowest magnetic shielding factor is 3950, we attribute $\sigma_B$ to the instability of our current supplies. Indeed, it is compatible with the measured relative current stability \cite{reinhard2009design}. 
The atom temperature fluctuations are small compared to a typical experiment using evaporative cooling. This may again be due to the exceptional magnetic field stability, since the atom temperature is determined by the magnetic field at the trap bottom during evaporation and the subsequent opening of the magnetic trap. At all stages, the current control is the most crucial. 
Using equations \ref{sigmaB} and \ref{sigmaT}, the temperature and magnetic field fluctuations translate into a frequency noise of $\sigma_{y,T}=1.0\times10^{-13}$ and $\sigma_{y,B}=0.7\times10^{-13}$ shot-to-shot, respectively.  The comparison in table \ref{noise_budget} shows, that these  are the main sources of frequency instability together with the Dick effect. Therefore, improving the magnetic field and temperature noise is of paramount importance. 
The atom temperature can in principle be extracted from the absorption images, which we take at each shot. Analysis of the data set of figure \ref{fig:TACC_stability_shots}
 gives shot-to-shot fluctuations of $\sigma_T/T=2-4\%$, which is much bigger than the 0.55\% deduced above. We therefore conclude that the determination of the cloud width is overshadowed by a significant statistical error. Nevertheless, it needs to be investigated, whether better detection and/or imaging at long time-of-flight, may reduce this error.
The magnetic field stability may be improved by refined power supplies, the use of multi-wire traps \cite{esteve2005realizing}, microwave dressing \cite{sarkany2014controlling} or ultimately the use of atom chips with permanent magnetic material \cite{sinclair2005bose,jose2014periodic,leung2014magnetic}.
If the magnetic field fluctuations  can be reduced, the temperature fluctuations may also reduce. Small $\sigma_B$ would also allow to operate nearer to  $\delta B_0^T$. 

\section{Conclusion}
We have built and characterised a compact atomic clock using magnetically trapped atoms on an atom chip. The clock stability reaches $5.8\times 10^{-13}$ at 1~s and is likely to integrate into the $10^{-15}$ range in less than a day. This is similar to the performance of the best compact atomic microwave clocks under development. 
It furthermore demonstrates the high degree of technical control that can be reached with atom chip experiments.
After correction for atom number fluctuations, variations of the atom temperature and magnetic field are the dominant causes of the clock instability together with the local oscillator noise. The magnetic field stability may be improved by additional current sensing and feedback and ultimately by the use of permanent magnetic materials. This would allow to operate nearer to the second sweet spot where the clock frequency is independent from temperature fluctuations.
The local oscillator noise takes an important role, because the clock duty cycle is $<30$\%. 
We are now in the process of designing a second version of this clock, incorporating fast atom loading and non-destructive atom detection. We thereby expect to reduce several  noise contributions to below $1\times10^{-13}\tau^{-1/2}$. 

\begin{acknowledgements}
We acknowledge fruitful discussion with P. Wolf, C. Texier and P. Uhrich. We thank M. Abgrall and J. Gu\'ena for the operation of the maser and the CSO. This work was supported by the EU under the project EMRP IND14.
\end{acknowledgements}

\bibliographystyle{apsrev4-1}
\bibliography{ClockStability,bibliography}



\end{document}